\documentclass[aps,prd,twocolumn,superscriptaddress,nofootinbib,longbibliography]{revtex4-2}

\usepackage{amsmath,amssymb}
\usepackage{graphicx}
\usepackage{bm}
\usepackage[colorlinks=true,linkcolor=blue,citecolor=blue,urlcolor=blue]{hyperref}
\usepackage{orcidlink}

\begin{document}

\title{Entropy production during reheating after inflation and \\
General Relativistic Entropic Acceleration}

\author{Juan Garc\'ia-Bellido \orcidlink{0000-0002-9370-8360}}
\email[]{juan.garciabellido@uam.es}
\affiliation{Instituto de F\'isica Te\'orica UAM-CSIC, Universidad Auton\'oma de Madrid, Cantoblanco 28049 Madrid, Spain}

\date{\today}

\begin{abstract}
Preheating transforms the coherent, low-entropy energy of the post-inflationary inflaton
condensate into a far-from-equilibrium bath of field quanta, and is therefore an
intrinsically irreversible process. We quantify that irreversibility directly, by computing
the coarse-grained phase-space entropy of the field fluctuations from classical-statistical
lattice simulations performed with \textsc{CosmoLattice}. For each field we reconstruct the
mode occupation numbers $n_k$ and evaluate $S=\sum_{\bm k}[(1+n_k)\ln(1+n_k)-n_k\ln n_k]$,
following its growth through the different stages of reheating in a self-consistently
expanding Friedmann--Lema\^itre--Robertson--Walker background. We study three canonical
models coupled to a daughter scalar $\chi$ through $g^2\phi^2\chi^2$. In the quadratic
$m^2\phi^2$ model ($q_{\rm res}=50$) broad parametric resonance drives the daughter
occupation numbers to $n_k\simeq7.5\times10^{2}$ in a narrow infrared band; the produced
entropy $\Delta S$ grows by a factor $\simeq2.5$, resides almost entirely in the daughter
sector, and saturates once backreaction shuts off the resonance. In the quartic
$\lambda\phi^4$ model ($q=g^2/\lambda=100$) the dynamics is radiation-like and strongly
nonlinear: occupation numbers cascade to the infrared and exceed $n_k\sim10^{11}$, the
produced entropy grows by more than three orders of magnitude to $\Delta S\simeq4\times10^6$
per comoving box, and rescattering shares it almost equally between the inflaton and daughter
fields. Finally, for symmetry-breaking tachyonic preheating we show that the entropy can be
obtained analytically: the spinodal growth of the long-wavelength modes makes the
Higgs occupation numbers grow exponentially, so that the entropy grows linearly in time,
$S_\phi\simeq(m^4/16\pi)Vt$, until it saturates at symmetry breaking. We discuss the ultraviolet vacuum floor intrinsic to the lattice initial conditions and argue that the produced entropy $\Delta S$ 
is the physically robust diagnostic. The phase-space entropy thus
provides a compact, dynamical measure that cleanly separates the resonance-dominated,
turbulence-dominated, and spinodal regimes of preheating. Finally, we couple this entropy production back into the expansion through the general
relativistic entropic acceleration (GREA) framework, modifying the self-consistent scale-factor
evolution of \textsc{CosmoLattice} to include the entropic force. The resulting acceleration
tracks the \emph{heat} (the energy transferred to particles) rather than the entropy itself:
it is substantial in the tachyonic and quartic-turbulent cases, where an $\mathcal{O}(1)$
fraction of the coherent energy fragments into radiation, but negligible in the resonant
$m^2\phi^2$ model despite its copious entropy production. The slow perturbative reheating that
follows generates the bulk of the cosmic entropy ($S_0 \sim10^{88}$) yet, being quasi-adiabatic,
drives no appreciable acceleration and shuts off entirely at thermalization.
\end{abstract}

\maketitle

\section{Introduction}
\label{sec:intro}

Reheating is the causal bridge between the end of inflation and the hot big bang: the energy
stored in the inflaton condensate must be transferred to the relativistic degrees of freedom
that dominate the radiation era, thermalizing the universe and setting the stage for big-bang
nucleosynthesis~\cite{Kofman:1994rk,Kofman:1997yn,Allahverdi:2010xz,Amin:2014eta}. In a broad
class of models this transfer begins explosively, through \emph{preheating}: the coherent
oscillations of the inflaton parametrically amplify the fluctuations of fields coupled to it,
or, after hybrid inflation~\cite{Garcia-Bellido:1997hex}, a tachyonic instability of the 
symmetry-breaking field converts the vacuum energy into classical waves within a single oscillation~\cite{Felder:2000hj}, populating a
far-from-equilibrium spectrum of modes long before perturbative decay or thermalization can
act~\cite{Traschen:1990sw,GarciaBellido:2001cb,GarciaBellido:2002aj,Copeland:2002ku}.
Preheating is inherently nonlinear: the amplified fluctuations backreact on the condensate,
rescatter off one another, and drive the system through a turbulent stage toward
equilibrium~\cite{Khlebnikov:1996mc,Micha:2004bv,Micha:2002ey}, so its quantitative study
requires lattice simulations that capture the full classical field dynamics in an expanding
universe.

While the energetics, the field spectra, and the associated gravitational-wave and relic
signatures of preheating have been studied in great detail~\cite{Khlebnikov:1997di,Garcia-Bellido:2007nns,Garcia-Bellido:2007fiu,Diaz-Gil:2007fch}, the \emph{entropy} generated as
the system evolves from a coherent condensate to a broad distribution of quanta is a
comparatively underused diagnostic. Yet entropy is precisely the quantity that captures the
irreversibility of the process: it distinguishes a single macroscopically occupied mode
(the condensate) from the same energy spread over many modes, and it grows
monotonically as phase space is explored. Because the occupation numbers reached during
preheating are large, $n_k\gg1$, the field dynamics is accurately classical, and a natural,
gauge-invariant measure of irreversibility is the coarse-grained phase-space entropy of the
mode distribution. Tracking this quantity through the different stages of preheating provides
a physically transparent, single-number summary of how far the system has evolved from its
initial coherent state, complementary to the usual spectral and energetic analyses. Moreover,
whenever the relevant mode functions are known in closed form, as they are in the linear
spinodal stage of tachyonic preheating, the entropy itself can be computed analytically,
providing a benchmark against which the lattice results can be tested.

In this work we compute the phase-space entropy directly from nonlinear lattice simulations
performed with \textsc{CosmoLattice}~\cite{Figueroa:2020rrl,Figueroa:2021yhd}, the modern 
C\texttt{++} successor of the pioneering \textsc{LatticeEasy} code~\cite{Felder:2000hq}, a
C\texttt{++} framework for lattice simulations of interacting scalar and gauge fields in an
expanding universe. We reconstruct the occupation numbers of each field on the lattice and
build the entropy shell by shell in momentum space. To expose how the entropy production
depends on the character of the reheating dynamics, we study three canonical single-field
models coupled to a daughter scalar: the quadratic $m^2\phi^2$ model, whose oscillations are
harmonic and whose background is matter-like; the quartic $\lambda\phi^4$ model, whose
oscillations are anharmonic and whose background is radiation-like and nearly conformal; and
the symmetry-breaking model $V=\tfrac14\lambda(\phi^2-v^2)^2$, in which the field rolls off an
unstable maximum and undergoes tachyonic (spinodal) preheating. These three cases span much of
the phenomenology of single-field reheating and, as we show, produce qualitatively different
entropy histories: a modest, daughter-dominated, resonance-limited growth in the quadratic
case; an explosive, democratically shared, turbulence-dominated growth in the quartic case;
and an analytically tractable, linear-in-time growth in the tachyonic case. We then feed this 
entropy production back into the cosmic expansion through the general relativistic
entropic acceleration (GREA) framework~\cite{EspinosaPortales:2021cac,GarciaBellido:2021jdr},
modifying the scale-factor evolution of \textsc{CosmoLattice} to include the entropic force, and
find that the induced acceleration is controlled by the \emph{energy} transferred to
particles: sizable for the tachyonic and turbulent transitions, negligible for resonant
$m^2\phi^2$, and absent once the subsequent perturbative reheating thermalizes and freezes the
entropy.

The remainder of the paper is organized as follows. Section~\ref{sec:entropy} introduces the
coarse-grained phase-space entropy and its lattice implementation. Sections~\ref{sec:mphi2},
\ref{sec:lphi4}, and \ref{sec:ssb} present the quadratic, quartic, and symmetry-breaking
models, respectively, comparing the lattice results with analytical expectations and, in the
tachyonic case, with a closed-form entropy law. Section~\ref{sec:discussion} compares the
three, discusses the ultraviolet vacuum floor and the role of the classical approximation, and
identifies the robust features of the result. We conclude in Sec.~\ref{sec:conclusions}. Appendix~\ref{app:code} documents the modifications to \textsc{CosmoLattice} implementing the three models, the occupation-number output, and the GREA entropic force.

\section{Phase-space entropy}
\label{sec:entropy}

We consider two-field models with action
\begin{equation}
S=\int d^4x\,\sqrt{-g}\left[-\tfrac12(\partial\phi)^2-\tfrac12(\partial\chi)^2-V(\phi,\chi)\right],
\end{equation}
in a spatially flat FLRW background whose expansion is sourced self-consistently by the
fields, with $\phi$ the inflaton and $\chi$ a daughter scalar coupled through
$g^2\phi^2\chi^2$. \textsc{CosmoLattice} integrates the discretized equations of motion in
dimensionless ``program'' variables, rescaling the fields by a characteristic amplitude
$f_\star$ and spacetime by a characteristic frequency $\omega_\star$, with a scale-factor
rescaling exponent $\alpha$ chosen according to the background; the models differ only in
these rescalings and in the potential, specified below.

For a nearly Gaussian, statistically homogeneous and isotropic field, the reduced density
matrix of each Fourier mode is characterized by its occupation number $n_k$. Tracing out the
phases yields the coarse-grained von Neumann (phase-space) entropy ($k_B=1$) of a bosonic
mode,
\begin{equation}
s(n_k)=(1+n_k)\ln(1+n_k)-n_k\ln n_k ,
\label{eq:smode}
\end{equation}
which vanishes for an empty mode and grows as $s\simeq\ln n_k$ for $n_k\gg1$; this coarse-grained 
entropy of an amplified stochastic field was introduced for cosmological perturbations in Refs.~\cite{Brandenberger:1992jh,Gasperini:1993yf}. Summing over
the lattice momenta gives the total entropy carried by the fluctuations of a field in the
comoving simulation volume,
\begin{equation}
S(t)=\sum_{\bm k}s\big(n_k(t)\big)
     =\sum_{\rm bins}\mathcal{N}_{\rm bin}\,s\big(n_{\rm bin}(t)\big),
\label{eq:Stot}
\end{equation}
with $\mathcal{N}_{\rm bin}$ the number of lattice modes in each spherical shell of the
radially binned spectrum. The occupation number is reconstructed on the lattice from the field
and its conjugate momentum; for the parametric models we use the standard estimator
$n_k=\tfrac{1}{2\omega_k}\langle\omega_k^2|\tilde\phi_{\bm k}|^2+|\tilde\pi_{\bm k}|^2\rangle$
with $\omega_k=\sqrt{k^2+a^2m_{\rm eff}^2}$, while for the tachyonic model, where $m_{\rm eff}^2$
is transiently negative, we use the frequency-independent definition
$n_k+\tfrac12=|\tilde\phi_{\bm k}\tilde\pi_{\bm k}|=\sqrt{P_\phi P_\pi}$ (Sec.~\ref{sec:ssb}).
We report both the total $S$ and the produced entropy $\Delta S(t)=S(t)-S(0)$, which subtracts
the static contribution of the initial vacuum fluctuations.

\section{Quadratic $m^2\phi^2$ preheating}
\label{sec:mphi2}

\subsection{Model and setup}

The quadratic model has potential
\begin{equation}
V(\phi,\chi)=\tfrac12 m^2\phi^2+\tfrac12 g^2\phi^2\chi^2 ,
\label{eq:Vmphi2}
\end{equation}
so that after inflation the inflaton executes harmonic oscillations with amplitude $\phi_0$
about the quadratic minimum, driving a matter-like background. The linearized daughter
fluctuations obey a Mathieu equation whose instability structure is governed by
$q_{\rm res}=g^2\phi_0^2/(4m^2)$; for $q_{\rm res}\gg1$ the system is in the regime of broad
parametric resonance~\cite{Kofman:1997yn}. We rescale $\tilde\phi=\phi/f_\star$ with
$f_\star=\phi_0$, take $\omega_\star=m$ and $\alpha=0$, and implement Eq.~\eqref{eq:Vmphi2} as
a custom model. The program potential is
$\tilde V=\tfrac12\tilde\phi^2+\tfrac12 q\,\tilde\phi^2\tilde\chi^2$ with $q=4q_{\rm res}$.

We evolve the system on a lattice of $N^3=48^3$ points with the velocity-Verlet integrator and
self-consistent expansion. We take $m=10^{-6}M_{\rm Pl}$ (reduced Planck mass
$M_{\rm Pl}=2.435\times10^{18}\,$GeV), $\phi_0=M_{\rm Pl}$ with the slow-roll attractor
velocity, $q_{\rm res}=50$, and integrate to $mt=150$ with $m\Delta t=5\times10^{-3}$. The
scale factor grows by $a\simeq21$ ($\simeq3$ $e$-folds) and energy is conserved to
$\max|\Delta E/E|=1.2\times10^{-4}$.

\subsection{Results}

\begin{figure*}[t]
\centering
\includegraphics[width=0.48\textwidth]{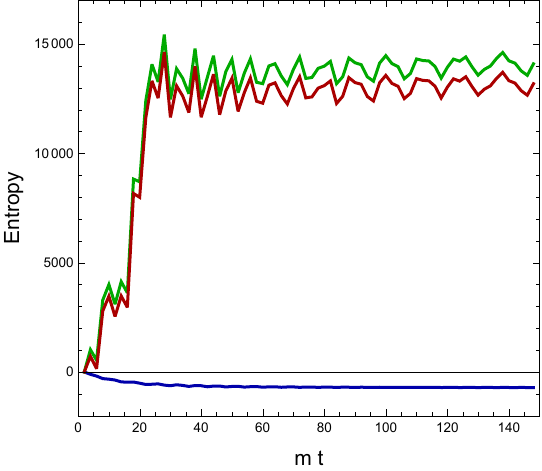}
\hspace{3mm}
\includegraphics[width=0.48\textwidth]{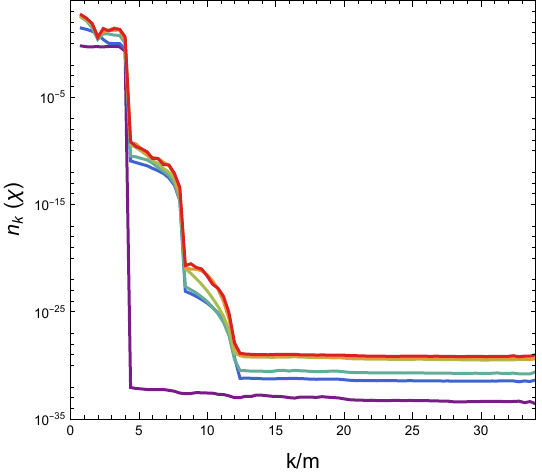}
\caption{Quadratic $m^2\phi^2$ model ($q_{\rm res}=50$), $N^3=100^3$, followed to
$mt\simeq150$. \emph{Left:} produced phase-space entropy $\Delta S(t)=S(t)-S(0)$ in the comoving
box (red), decomposed into the daughter $\chi$ (green) and inflaton $\phi$ (blue); the entropy
rises through the resonance and then saturates. The inflaton curve is slightly negative---a
vacuum-subtraction artifact discussed in the text, not physical entropy loss. \emph{Right:} the
daughter occupation-number spectrum $n_k$ at a sequence of times, showing the growth of the
narrow infrared resonance band (peak $n_k\simeq7\times10^{2}$ at $k/m\simeq0.8$) and its
subsequent rescattering-driven broadening.}
\label{fig:mphi2}
\end{figure*}

Figure~\ref{fig:mphi2} summarizes the run. Starting from the vacuum level $n_k\lesssim0.5$,
the daughter modes in a narrow infrared band grow exponentially: by $mt\simeq22$ the band
peaks at $n_k\simeq7.5\times10^{2}$ around $k/m\simeq0.8$, the hallmark of broad parametric
resonance. As $\langle\chi^2\rangle$ grows, its backreaction on the inflaton (through the
effective mass $g^2\langle\chi^2\rangle$) detunes the resonance, and the spectrum saturates.
Correspondingly, the produced entropy climbs steeply during $10\lesssim mt\lesssim24$ and
then \emph{saturates}: once the backreaction shuts off the resonance, $\Delta S$ plateaus,
rising only slowly thereafter (by a further $\sim10\%$ out to $mt\simeq100$) as rescattering
broadens the daughter spectrum toward the ultraviolet (right panel of Fig.~\ref{fig:mphi2}).
Quantitatively, the total entropy grows from $S(0)\simeq9.0\times10^3$ to a plateau
$S\simeq2.2\times10^4$, a produced entropy $\Delta S\simeq1.3\times10^4$ (a factor $\simeq2.4$
increase), essentially all of which resides in the daughter sector.

The inflaton's own contribution, by contrast, is slightly \emph{negative},
$\Delta S_\phi\simeq-8\times10^{2}$ ($-6\%$ of the total). This is not a violation of the
second law but an instructive artifact of the occupation-number definition. In $m^2\phi^2$
preheating the inflaton itself is not parametrically amplified (only the daughter is resonant) 
so its genuine particle production, and hence its physical entropy production, is
zero. The lattice occupation number is nonetheless defined relative to the instantaneous
adiabatic vacuum, $n_k=(2\omega_k)^{-1}\langle\omega_k^2|\phi_k|^2+|\pi_k|^2\rangle$ with
$\omega_k=\sqrt{k^2+a^2 m_{\rm eff}^2}$. The seeded initial fluctuations register
$n_k\simeq0.5$--$0.7$, slightly \emph{above} the instantaneous vacuum; as the universe expands
($a$ grows by $\sim7$) and the inflaton effective mass tracks the growing
$g^2\langle\chi^2\rangle$ backreaction, these modes relax toward the adiabatic vacuum
$n_k\to\tfrac12$, so the reconstructed occupation, and with it the summed single-mode
entropy, drifts slightly \emph{down}. We have verified that the effect is concentrated
entirely in the ultraviolet, where the many near-vacuum modes live
($\Delta S_\phi^{\rm UV}\simeq-8\times10^{2}$ versus $\Delta S_\phi^{\rm IR}\simeq-10$), and
that it is essentially unchanged across $N^3=48^3$, $100^3$, and $128^3$. It should therefore be read as
the residual ambiguity of the vacuum subtraction in a sector with no real particle production,
confirming that all the physical entropy is generated in the daughter field; a strict
adiabatic (Bogoliubov) vacuum subtraction would set $\Delta S_\phi\equiv0$.

\section{Quartic $\lambda\phi^4$ preheating}
\label{sec:lphi4}

\subsection{Model and setup}

The quartic model has potential
\begin{equation}
V(\phi,\chi)=\tfrac14\lambda\phi^4+\tfrac12 g^2\phi^2\chi^2 ,
\label{eq:Vlphi4}
\end{equation}
the built-in \texttt{lphi4} model of \textsc{CosmoLattice}. For the quartic minimum the
inflaton oscillations are anharmonic, the background equation of state is radiation-like,
$\langle w\rangle\simeq1/3$, and the dynamics is nearly conformal. The resonance is
controlled by $q=g^2/\lambda$. We rescale $\tilde\phi=\phi/f_\star$ with $f_\star=\phi_0$, take
$\omega_\star=\sqrt{\lambda}\,\phi_0$ and $\alpha=1$, so that the natural time and momentum
variables are $\sqrt{\lambda}\,\phi_0\,t$ and $\kappa=k/(\sqrt{\lambda}\,\phi_0)$; the program
potential is $\tilde V=\tfrac14\tilde\phi^4+\tfrac12 q\,\tilde\phi^2\tilde\chi^2$.

We evolve the system on a lattice of $N^3=48^3$ points with self-consistent expansion, with
$\lambda=9\times10^{-14}$, $\phi_0=5.70\times10^{18}\,$GeV and the inflationary slow-roll
velocity, and $q=100$, integrating to $\sqrt{\lambda}\phi_0 t=300$ with
$\sqrt{\lambda}\phi_0\Delta t=10^{-2}$. The scale factor grows by $a\simeq2.6\times10^{2}$
($\simeq5.6$ $e$-folds) and energy is conserved to $\max|\Delta E/E|=8\times10^{-4}$.

\subsection{Results}

\begin{figure*}[t]
\centering
\includegraphics[width=0.48\textwidth]{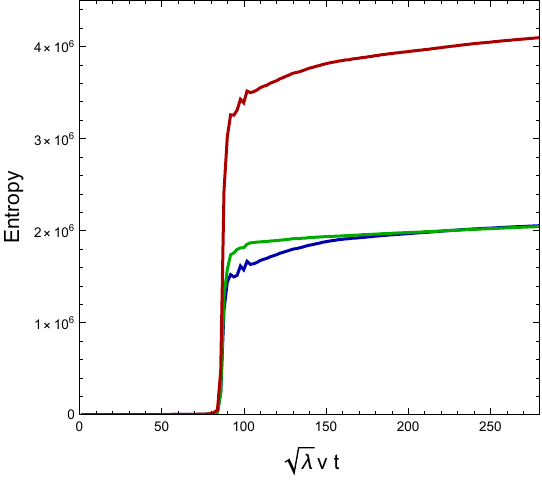}
\hspace{3mm}
\includegraphics[width=0.48\textwidth]{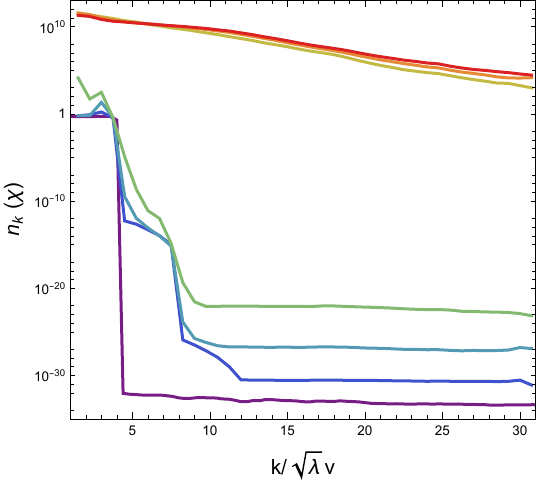}
\caption{Quartic $\lambda\phi^4$ model ($q=g^2/\lambda=100$). \emph{Left:} produced
phase-space entropy $\Delta S(t)$ in the comoving box (red), with the daughter $\chi$ (green)
and inflaton $\phi$ (blue) contributions, versus $\sqrt{\lambda}\,v\,t$. \emph{Right:} the
daughter occupation-number spectrum $n_k$ at a sequence of times, showing resonant growth
followed by the infrared cascade characteristic of wave turbulence.}
\label{fig:lphi4}
\end{figure*}

Figure~\ref{fig:lphi4} summarizes the run. The daughter modes grow exponentially from the
vacuum; by $\sqrt{\lambda}\phi_0 t\simeq80$ the occupation numbers already span many orders of
magnitude, and as the fields become nonlinear the spectrum cascades toward the infrared, with
the lowest bins exceeding $n_k\sim10^{11}$. This infrared condensation is the lattice
signature of the relativistic wave turbulence that carries the system slowly toward
equilibrium~\cite{Micha:2004bv,Micha:2002ey,Podolsky:2005bw}. Rescattering feeds power back into the inflaton
fluctuations, whose occupation numbers grow to values comparable to those of the daughter
field. The total entropy grows from a small vacuum value $S(0)\simeq1.3\times10^3$ to
$S\simeq4.1\times10^6$, a produced entropy $\Delta S\simeq4.1\times10^6$, an increase by a
factor $\simeq3.3\times10^3$. The entropy is shared almost equally between the two sectors
($50\%$ each), reflecting efficient rescattering, and continues to rise well into the
nonlinear stage, reaching $90\%$ of its final value only around
$\sqrt{\lambda}\phi_0 t\simeq130$ before saturating.

\section{Tachyonic preheating from symmetry breaking}
\label{sec:ssb}

\subsection{Model and setup}

The symmetry-breaking model has potential
\begin{equation}
V(\phi,\chi)=\tfrac14\lambda\,(\phi^2-v^2)^2+\tfrac12 g^2\phi^2\chi^2 ,
\label{eq:Vssb}
\end{equation}
studied in Ref.~\cite{GarciaBellido:2001cb} as a model of tachyonic (spinodal) preheating
after hybrid inflation~\cite{Garcia-Bellido:1997hex,Felder:2000hj,Felder:2001kt,GarciaBellido:2002aj,Copeland:2002ku}. The Higgs field starts at the top
of the potential, $\phi=0$, with only vacuum fluctuations. The curvature there is negative,
$V''(0)=-m^2$ with $m^2=\lambda v^2$, so all long-wavelength modes $k<m$ are tachyonic and
grow spinodally. We rescale $\tilde\phi=\phi/v$ with $f_\star=v$, take $\omega_\star=m$ and
$\alpha=0$, and switch off the expansion ($a=1$), appropriate to the fast quench; the program
potential is $\tilde V=\tfrac14(\tilde\phi^2-1)^2+\tfrac12 q\,\tilde\phi^2\tilde\chi^2$ with
$q=g^2/\lambda\equiv\alpha^2$. Since $V''(0)<0$, \textsc{CosmoLattice} automatically seeds the
Higgs fluctuations with the massless vacuum dispersion $\omega_k=|k|$, i.e.
$\phi_k(0)=1/\sqrt{2k}$, matching exactly the initial conditions of
Ref.~\cite{GarciaBellido:2001cb}. We use $N^3=128^3$ with $k_{\rm IR}=0.02\,m$ (box length
$L=100\pi/m$, $k_{\rm max}=2.22\,m$, exactly as in Ref.~\cite{GarciaBellido:2001cb}), with
$\lambda=10^{-4}$, $q=2$, and integrate to $mt=30$ with $m\Delta t=4\times10^{-2}$; energy is conserved to
$\max|\Delta E/E|=9\times10^{-5}$.

Because the linear (spinodal) stage admits closed-form mode functions, the phase-space entropy
of the Higgs can be obtained analytically. With vacuum initial conditions the tachyonic modes
grow as $\phi_k(t)=\phi_k(0)\,e^{t\sqrt{m^2-k^2}}$, and with the frequency-independent
definition $n_k+\tfrac12=|\phi_k^*\dot\phi_k|$ of Ref.~\cite{GarciaBellido:2001cb} the
occupation numbers are
\begin{equation}
n_k(t)+\tfrac12=\tfrac12\,e^{2t\sqrt{m^2-k^2}},
\label{eq:nk}
\end{equation}
which are large, $n_k\gg1$, throughout the growing band $k<m$. The single-mode
entropy~\eqref{eq:smode} then reduces to $s(n_k)\simeq\ln n_k+1\simeq2t\sqrt{m^2-k^2}+1-\ln2$,
and the total Higgs entropy per unit volume follows by quadrature,
\begin{align}
\frac{S_\phi(t)}{V}
&=\frac{1}{2\pi^2}\int_0^m\! dk\,k^2\left[2t\sqrt{m^2-k^2}+1-\ln2\right]\nonumber\\
&=\frac{m^4}{16\pi}\,t+\mathcal{O}(m^3),
\label{eq:Slinear}
\end{align}
where we used $\int_0^m k^2\sqrt{m^2-k^2}\,dk=\pi m^4/16$. The entropy therefore grows
\emph{linearly} in time during tachyonic preheating, a direct consequence of the occupation
numbers growing exponentially, $\ln n_k\propto t$, at a constant rate
$dS_\phi/dt\simeq(m^4/16\pi)V$. The linear rise continues until the symmetry breaks at
$t\simeq t_*$, when $\langle\phi^2\rangle\to v^2$ and the tachyonic growth terminates;
matching to the peak occupation number at symmetry breaking,
$n_k(t_*)=(16\pi/\lambda)\,e^{-k^2/2k_*^2}$~\cite{GarciaBellido:2001cb}, fixes
$mt_*\simeq\tfrac12\ln(32\pi^2/\lambda)$. Thereafter the occupation numbers freeze and the
entropy saturates at
\begin{equation}
S_\phi^{\rm max}\simeq\frac{m^4}{16\pi}\,V\,t_*\simeq\frac{m^3V}{32\pi}\,\ln\!\frac{32\pi^2}{\lambda}.
\label{eq:Smax}
\end{equation}
The daughter entropy is instead controlled by the time-independent Bogoliubov occupation
number $n_k^B(\alpha)$ of Ref.~\cite{GarciaBellido:2001cb}, with $\alpha=g/\sqrt{\lambda}$, and
is a constant $S_\chi=V\!\int\!d^3k/(2\pi)^3\,s(n_k^B)$, small for the weak couplings
considered here. 

\subsection{Results}

\begin{figure*}[t]
\centering
\includegraphics[width=0.325\textwidth]{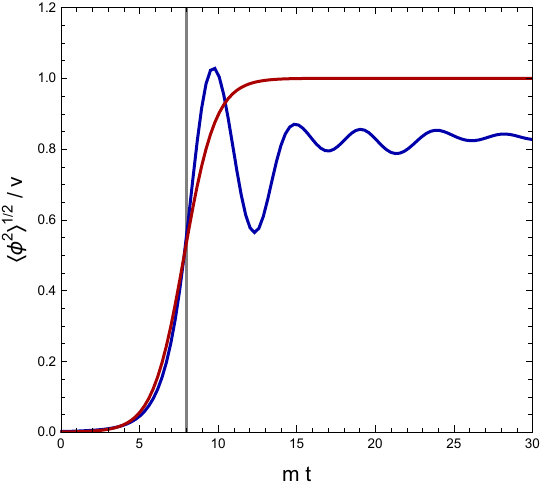}
\includegraphics[width=0.325\textwidth]{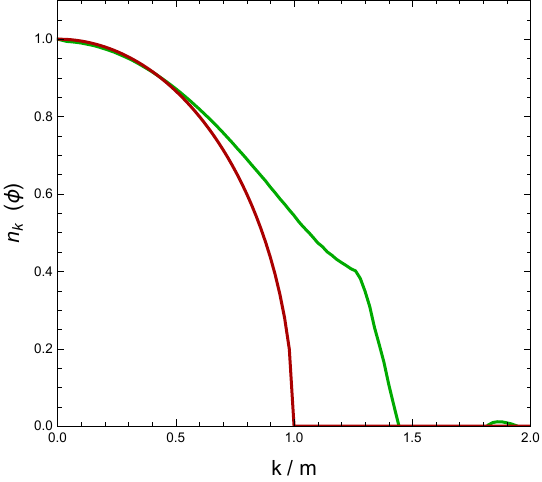}
\includegraphics[width=0.325\textwidth]{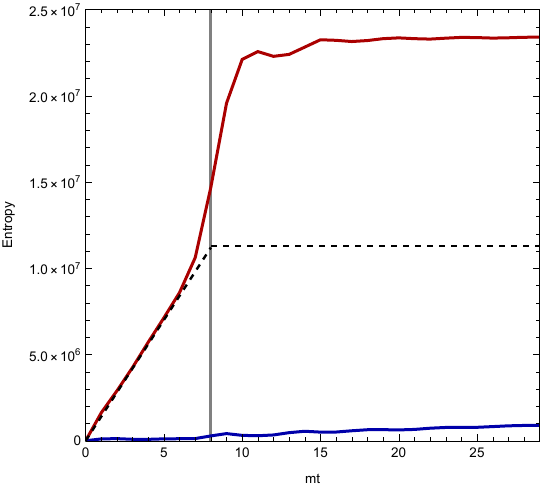}
\caption{Symmetry-breaking (tachyonic) model of Ref.~\cite{GarciaBellido:2001cb}, simulated
with \textsc{CosmoLattice}. \emph{Left:} the lattice-averaged $\langle\phi^2\rangle^{1/2}/v$
(blue) follows the tanh profile of Eq.~(6) of Ref.~\cite{GarciaBellido:2001cb} (red), with a
symmetry-breaking time $mt_*\simeq7.9$ (gray), then oscillates about the true vacuum as in Fig.~1 of Ref.~\cite{GarciaBellido:2001cb}. \emph{Center:} the measured spinodal growth rate
$\gamma_k=\tfrac12\,d\ln P_\phi/dt$ (green) reproduces the tachyonic dispersion relation
$\sqrt{m^2-k^2}$ of Eq.~\eqref{eq:nk} (red) across the unstable band. \emph{Right:} the
phase-space entropy: the Higgs entropy $S_\phi$ (red) grows linearly during the spinodal stage
and turns over at $t\simeq t_*$, then saturates to a constant plateau, in agreement with the analytic law $\propto(m^4/16\pi)V\min(t,t_*)$
(dashed); the daughter entropy $S_\chi$ (blue) remains small.}
\label{fig:ssb}
\end{figure*}

Figure~\ref{fig:ssb} compares the lattice evolution with these analytical expressions. The
left panel shows the lattice-averaged $\langle\phi^2\rangle^{1/2}/v$, which follows the tanh
profile $\phi(t)=\tfrac{v}{2}[1+\tanh(m(t-t_*)/2)]$ of Ref.~\cite{GarciaBellido:2001cb} with a
symmetry-breaking time $mt_*\simeq7.9$, close to the value $mt_*=8$ of
Ref.~\cite{GarciaBellido:2001cb}, followed by the weakly damped oscillations about the true
vacuum. The central panel shows the measured spinodal growth rate
$\gamma_k=\tfrac12\,d\ln P_\phi/dt$, which reproduces the tachyonic dispersion relation
$\sqrt{m^2-k^2}$ of Eq.~\eqref{eq:nk} across the unstable band, with a correlation coefficient
of $0.98$ and a maximum rate $\gamma_0\simeq m$. The right panel
shows the phase-space entropy: reconstructing $n_k$ on the lattice from the field and momentum
spectra via $n_k+\tfrac12=|\phi_k\dot\phi_k|=\sqrt{P_\phi P_\pi}$, self-calibrated on the
initial vacuum (equivalent to the regularized dispersion of
Ref.~\cite{GarciaBellido:2001cb}), the Higgs entropy $S_\phi$ grows linearly during the
spinodal stage and turns over at $t\simeq t_*$, in agreement with
Eqs.~\eqref{eq:Slinear}--\eqref{eq:Smax}; the measured slope agrees with the leading-order
estimate $(m^4/16\pi)V$ to within a factor of order unity. For $t>t_*$ the entropy saturates to a constant plateau (the post-transition Higgs oscillations are adiabatic and drive no further particle production) confirming that entropy generation ceases once symmetry breaking is complete. The daughter entropy $S_\chi$
remains small, consistent with the weak coupling and with the negligible backreaction noted in
Ref.~\cite{GarciaBellido:2001cb}.

\section{GREA implementation: entropic acceleration from preheating}
\label{sec:grea}

The entropy computed above is not merely a diagnostic. In the general relativistic
entropic acceleration (GREA) framework~\cite{EspinosaPortales:2021cac,GarciaBellido:2021jdr},
the growth of entropy in out-of-equilibrium epochs, such as (p)reheating, sources
a new entropic force on the expansion of the universe. Starting from the second law
$T\,dS=d(\rho a^3)+p\,d(a^3)$, which reduces to the usual continuity equation only when
$dS=0$, Ref.~\cite{GarciaBellido:2021jdr} obtains the modified system
\begin{align}
\dot\rho+3H(\rho+p)&=\frac{T\dot S}{a^3},\label{eq:grea1}\\
\dot a^2+k&=\frac{8\pi G}{3}\,\rho\,a^2,\label{eq:grea2}\\
\frac{\ddot a}{a}&=-\frac{4\pi G}{3}(\rho+3p)+\frac{4\pi G}{3}\frac{T\dot S}{a^3H},\label{eq:grea3}
\end{align}
where the last term of Eq.~\eqref{eq:grea3} is an entropic cosmological force, positive for
entropy production, that tends to accelerate the expansion. Reheating is expected to induce
``a second burst of accelerated expansion before the local fundamental interactions drive the
fluid to thermodynamical equilibrium''~\cite{GarciaBellido:2021jdr}; here we evaluate it
directly on the lattice. We switch on the self-consistent expansion in all three models
(for the symmetry-breaking case, replacing the fixed-background approximation $a=1$).

\subsection{Temperature and heat from the lattice}

The entropic term requires the computation of the heat $T\dot S$. We assign each mode the 
bosonic temperature
\begin{equation}
T_k=\frac{\omega_k}{\ln(1+1/n_k)},
\label{eq:Tk}
\end{equation}
{\it i.e.} the temperature for which $n_k=[\exp(\omega_k/T_k)-1]^{-1}$. Since the single-mode
entropy~\eqref{eq:smode} obeys $ds_k/dn_k=\ln(1+1/n_k)$, the heat carried by each mode is
$T_k\,ds_k=\omega_k\,dn_k$, and the logarithms cancel: the total heat is simply the rate of
energy flowing into the particle spectrum,
\begin{equation}
T\dot S=\sum_{\bm k}T_k\,\dot s_k=\sum_{\bm k}\omega_k\,\dot n_k .
\label{eq:heat}
\end{equation}
Equivalently, and more robustly on the lattice, $T\dot S/a^3$ is the non-adiabatic energy
transfer into the incoherent (kinetic$+$gradient) sector,
$T\dot S/a^3=\dot\rho_{\rm part}+3H(\rho_{\rm part}+p_{\rm part})$, which we evaluate from the
volume-averaged energies. It is convenient to express the effect through the dimensionless
entropic ratio
\begin{equation}
\mathcal{R}(t)\equiv\frac{T\dot S/a^3}{H(\rho+3p)},
\label{eq:Rratio}
\end{equation}
in terms of which Eq.~\eqref{eq:grea3} and the unchanged constraint~\eqref{eq:grea2} give a
deceleration parameter
\begin{equation}
q_{\rm GREA}=-\frac{\ddot a\,a}{\dot a^2}=q_{\rm std}\,(1-\mathcal{R}),\qquad
q_{\rm std}=\frac{\rho+3p}{2\rho}.
\label{eq:qgrea}
\end{equation}
Whenever $\mathcal{R}>1$ the entropic force overwhelms the gravitational deceleration and the
expansion accelerates, $q_{\rm GREA}<0$.

\subsection{Modification of CosmoLattice}

We implement Eqs.~\eqref{eq:grea1}--\eqref{eq:grea3} directly in the self-consistent
scale-factor evolution of \textsc{CosmoLattice}. The scale-factor kernel, which integrates the
program-variable form of Eq.~\eqref{eq:grea3}, is augmented with the entropic term (for the
$\alpha=0$ rescaling):
\begin{verbatim}
if (model.greaOn) {
  auto H = model.aDotI / model.aI;
  standard +=
     pow(model.aI, 2*model.alpha+1)/3.0
     * pow<2>(model.fStar/Model::MPl)
     * 0.5 * model.TdSrate
     / (pow<3>(model.aI) * H);
}
\end{verbatim}
where \texttt{TdSrate}$=T\dot S$ is supplied to the model by the measurer, computed from the
occupation-number spectra via Eq.~\eqref{eq:heat} (operator-split at the measurement cadence).
The Hamiltonian constraint~\eqref{eq:grea2} is monitored but no longer exactly enforced, its
violation quantifying the entropic energy exchange.

\subsection{Results}

\begin{figure*}[t]
\centering
\includegraphics[width=0.325\textwidth]{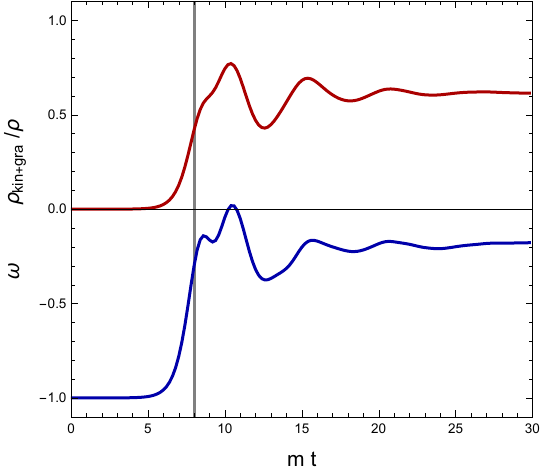}
\includegraphics[width=0.325\textwidth]{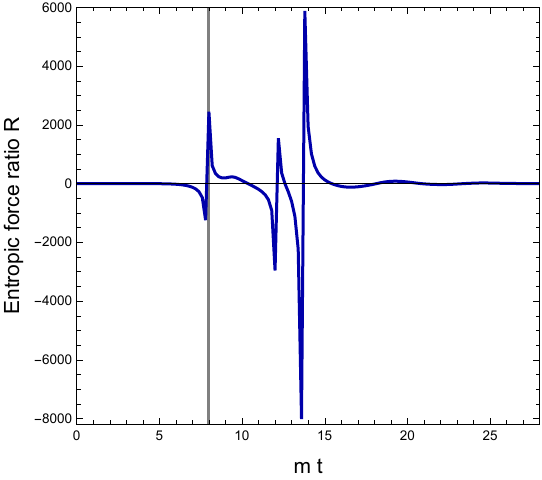}
\includegraphics[width=0.325\textwidth]{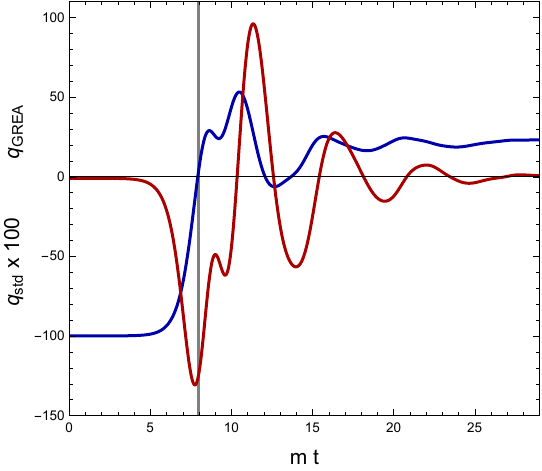}
\caption{GREA entropic acceleration during tachyonic preheating (symmetry-breaking model,
with expansion switched on). \emph{Left:} the equation of state $w$ (blue) evolves from the false
vacuum ($w=-1$) to the broken phase as the potential energy is converted into particles
($\rho_{\rm part}/\rho$, kinetic$+$gradient) (red). \emph{Center:} the entropic ratio
$\mathcal{R}$ of Eq.~\eqref{eq:Rratio} (signed-log scale) spikes by orders of magnitude during
the transition (shaded). \emph{Right:} the deceleration parameter; the standard
$q_{\rm std}\times100$ (blue) versus the GREA value $q_{\rm GREA}=q_{\rm std}(1-\mathcal{R})$ (red),
which plunges strongly negative during the burst of entropy production.}
\label{fig:grea}
\end{figure*}

The effect depends sharply on how much energy, not merely phase-space entropy, is actually
transferred. In the quadratic $m^2\phi^2$ model the coherent inflaton condensate retains
almost all of the energy ($\rho_{\rm part}/\rho\ll1$), so $\mathcal{R}\approx0$ and the
entropic force is negligible. The tachyonic symmetry-breaking model is the opposite extreme
and the natural setting for GREA: the false-vacuum energy is converted wholesale into
particles within about one oscillation. Figure~\ref{fig:grea} shows the result. Before the
transition the false vacuum drives an ordinary de Sitter phase ($w=-1$, $q_{\rm std}=-1$).
During the transition ($mt\simeq6$--$12$), the explosive entropy production makes $\mathcal{R}$
spike to $\mathcal{O}(10^3)$---amplified by the smallness of $H$ in the fast quench---and
$q_{\rm GREA}$ plunges strongly negative: a burst of entropic acceleration. Integrated over
the run, the universe accelerates ($q_{\rm GREA}<0$) $65\%$ of the time, against $32\%$ in the
standard case. This is a direct lattice realization of the ``second burst of accelerated
expansion'' anticipated in Ref.~\cite{GarciaBellido:2021jdr}, now sourced by the
first-principles entropy production of the preheating fields.

\subsection{Extra e-folds from the entropic burst}

The entropic force operates only while entropy is produced, {\it i.e.} during the tachyonic transition,
$mt\simeq6-12$; once the symmetry breaking completes and $\dot S\to0$
(Fig.~\ref{fig:ssb}), the source in Eq.~\eqref{eq:grea1} switches off and the expansion
returns to standard, decelerating evolution. Because the Hamiltonian
constraint~\eqref{eq:grea2} ties $H$ to the energy density, the effect is bounded: the burst
injects an entropic energy $\Delta\rho_S=\int(T\dot S/a^3)\,dt\simeq\tfrac14\lambda v^4$, of
order the released false-vacuum energy, which boosts $H$ by $\sim35\%$ and then redshifts away
as the universe expands.

\begin{figure*}[t]
\centering
\includegraphics[width=0.325\textwidth]{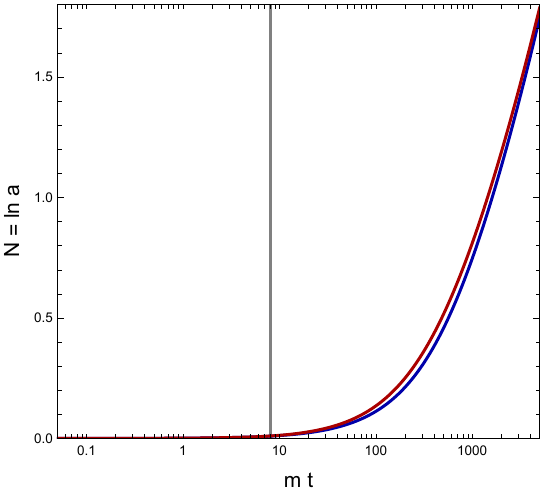}
\includegraphics[width=0.325\textwidth]{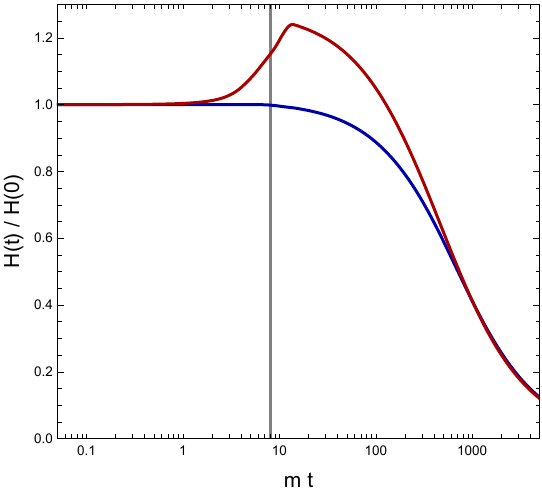}
\includegraphics[width=0.325\textwidth]{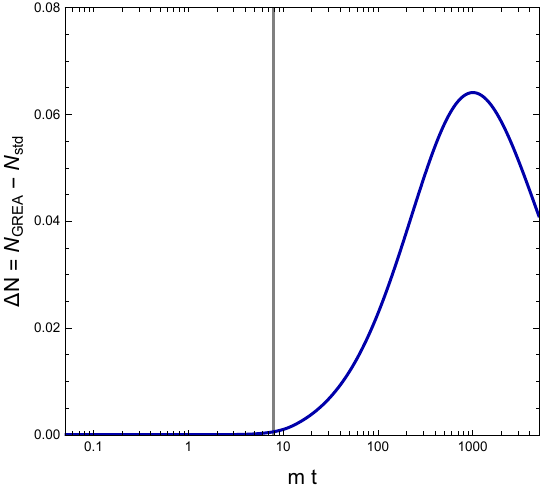}
\caption{Extra inflation from the entropic burst in the symmetry-breaking model, followed to
$mt=5000$ ($1/H\simeq840/m$). \emph{Left:} the e-folds $N=\ln a$; the background model
(standard, blue) reproduces the full nonlinear \textsc{CosmoLattice} evolution (green points, a
dedicated run to $mt=5000$) to better than $4\times10^{-4}$, validating the coupled
field$+$Friedmann integration, while the GREA branch (red) runs slightly ahead. \emph{Center:}
the Hubble rate is boosted by $\sim35\%$ during the burst (shaded) and then dilutes back to the
standard value by about one Hubble time. \emph{Right:} the extra e-folds
$\Delta N=N_{\rm GREA}-N_{\rm std}$ rise, peak at $\simeq0.06$ near one Hubble time, and then
decline as the faster-expanding GREA universe dilutes its own matter---a transient, bounded
burst rather than a sustained inflationary epoch.}
\label{fig:efolds}
\end{figure*}

To evaluate the effect we integrate Eqs.~\eqref{eq:grea1}--\eqref{eq:grea2} with the matter
equation of state taken from the lattice and the entropic energy, once injected, diluting as
radiation ($w_S=1/3$). Both branches are evolved self-consistently, each component redshifting
with its own branch's Hubble rate. We have verified the input against a dedi\-cated
\textsc{CosmoLattice} run of the symmetry-breaking model with self-consistent expansion carried
all the way to $mt=5000$: the equation of state settles to and oscillates about
$\langle w\rangle\simeq-0.2$ over the entire history (the $\mathbb{Z}_2$ domain-wall and
gradient network sustaining a mild negative pressure rather than relaxing to matter or
radiation) and the background model integrated with this $w(t)$ reproduces the measured
$N=\ln a$ to $<4\times10^{-4}$ (left panel of Fig.~\ref{fig:efolds}). The run itself spans less
than $4\%$ of a Hubble time at $mt=30$, which is why $\ln a$ appears linear there; followed to
$mt=5000$ the deceleration is manifest.

The resulting extra expansion is
\begin{equation}
\Delta N=N_{\rm GREA}-N_{\rm std}\simeq0.06 ,
\end{equation}
reaching a maximum $\simeq0.064$ around one Hubble time ($mt\simeq10^{3}$) before the boosted
$H$ relaxes and the excess slowly erodes. It is a transient, bounded burst of acceleration.
Since $\Delta\rho_S/\rho=\mathcal{O}(1)$ is set by the ratio of the entropic to the vacuum
energy rather than by the symmetry-breaking scale, this $\Delta N\sim0.06$ is largely
insensitive to $v$. It is a first-principles, lattice realization of the ``second burst of
accelerated expansion'' anticipated for reheating in Ref.~\cite{GarciaBellido:2021jdr}, and
would be more pronounced for transitions with a larger hierarchy between the latent heat and
the ambient energy density.

Two elements of the calculation remain approximate. The validation run uses a modest lattice
($N=32$), adequate for the background equation of state but not for detailed spectra; and while
the matter sector is now lattice-derived, the radiation-like dilution assumed for the entropic
energy, and the treatment of the entropic force as a background source rather than a per-step
feedback on the field dynamics, are modeling choices. A fully self-consistent coupling of the
entropic force to the lattice evolution is left to future work.

\section{Discussion}
\label{sec:discussion}

The three models produce qualitatively different entropy histories, and the contrast is
instructive. In the quadratic case the entropy production is limited by backreaction: the
resonance is narrow, it shuts off once $g^2\langle\chi^2\rangle$ detunes the inflaton
oscillations, and the produced entropy, confined almost entirely to the daughter, grows by
only a factor of a few before saturating. In the quartic case the near-conformal dynamics
sustains a much longer growth: the fields become strongly nonlinear, rescattering redistributes
power democratically, and an infrared cascade drives the occupation numbers to
$n_k\sim10^{11}-10^{12}$, so that the produced entropy grows by more than three orders of
magnitude. In the tachyonic case the entropy is generated by the spinodal instability rather
than by resonance and, since the linear mode functions are known, its history is
analytically calculable: the exponential growth of the occupation numbers, $\ln n_k\propto t$,
translates into a strictly \emph{linear} rise of the entropy, $S_\phi\simeq(m^4/16\pi)Vt$,
that saturates abruptly once the symmetry breaks. The phase-space entropy thus cleanly
separates a resonance-dominated regime, a turbulence-dominated regime, and a spinodal regime.

\begin{figure*}[t]
\centering
\includegraphics[width=0.485\textwidth]{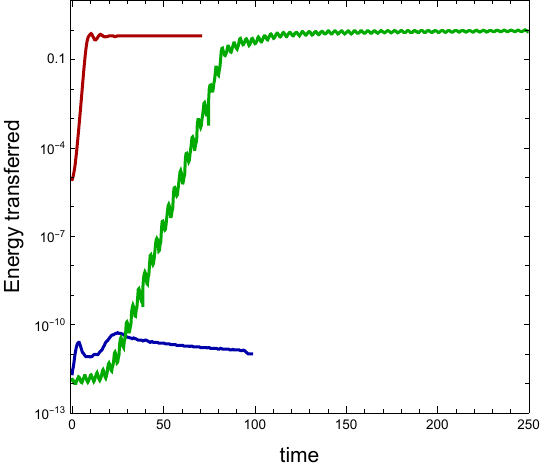}
\includegraphics[width=0.485\textwidth]{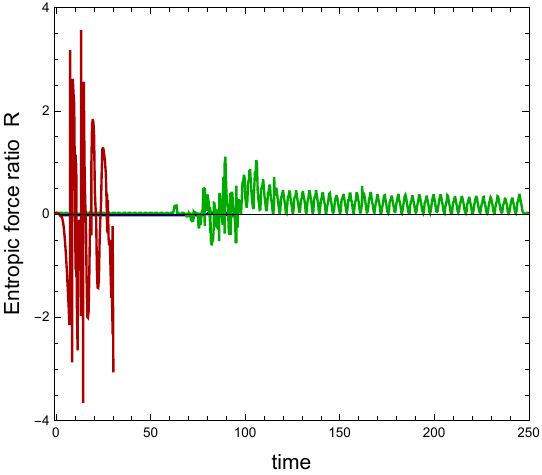}
\caption{The GREA entropic force across the three models, with self-consistent expansion.
\emph{Left:} fraction of energy transferred from the coherent field to fluctuations,
$\rho_{\rm fluc}/\rho$:\ negligible ($\sim10^{-11}$) for resonant $m^2\phi^2$ (blue), rising to $\sim0.9$
for the turbulent $\lambda\phi^4$ (green), and $\mathcal{O}(1)$ for the tachyonic symmetry-breaking
transition (red). \emph{Right:} the corresponding entropic ratio $\mathcal{R}$ (signed-log scale: sign$({\cal R})\,\log_{10}(1+|{\cal R})|)$,
spanning $|\mathcal{R}|\lesssim10^{-10}$, $|\mathcal{R}|\sim10$, and $|\mathcal{R}|\sim10^{3}$ for the resonant $m^2\phi^2$ (blue), the turbulent $\lambda\phi^4$ (green), and the tachyonic symmetry-breaking
model (red), respectively.
The entropic acceleration follows the energy transferred to particles, not the phase-space
entropy.}
\label{fig:grea3}
\end{figure*}

It is instructive to ask which ingredient of the entropy production actually sources the
entropic force of Sec.~\ref{sec:grea}. Applying the same GREA construction, with the
self-consistent expansion switched on, to all three models reveals that the force tracks the
\emph{heat} $T\dot S=\sum_k\omega_k\dot n_k$, {\it i.e.}~the \emph{energy} transferred into
particles, and not the phase-space entropy itself. The three cases span three regimes
(Fig.~\ref{fig:grea3}). In the quadratic $m^2\phi^2$ model the inflaton condensate survives: it
keeps oscillating and redshifting as pressureless matter, transferring only a
quantum-vacuum-level fraction $\rho_{\rm fluc}/\rho\sim10^{-11}$ into the daughter, so that
despite a large phase-space entropy ($\Delta S\sim10^{4}$) the entropic ratio never exceeds
$|\mathcal{R}|\sim10^{-10}$ and the induced expansion is $\Delta N\sim10^{-11}$, {\it i.e.} utterly
negligible. In the quartic $\lambda\phi^4$ model the inflaton instead \emph{fragments}: its
homogeneous amplitude decays and an $\mathcal{O}(1)$ fraction of the energy cascades into the
turbulent bath, $\rho_{\rm fluc}/\rho\to0.9$, sustaining $\mathcal{R}\sim10$ and a substantial
extra expansion, $\Delta N\sim\mathcal{O}(1)$. The tachyonic symmetry-breaking transition is the
extreme case, converting the false-vacuum energy wholesale into particles within a single
oscillation, $\mathcal{R}\sim10^{3}$, in a brief but intense burst. The GREA source being the
heat, the lesson is that the ``second burst'' of entropic acceleration is controlled by the
coherent energy actually dumped into the matter sector: both the quartic-turbulent and the
tachyonic transitions convert an $\mathcal{O}(1)$ fraction and drive appreciable acceleration,
whereas resonant $m^2\phi^2$ preheating, which produces copious entropy but transfers
negligible energy, does not.

An important caveat specific to classical-statistical lattice simulations concerns the initial
conditions, which seed every resolved mode with vacuum fluctuations, $n_k\sim\tfrac12$. The
entropy~\eqref{eq:Stot} therefore carries a nonzero, ultraviolet-dominated ``vacuum floor''
$S(0)$ already at $t=0$; this floor is sensitive to the lattice cutoff and does not correspond
to physical particle production, being the classical counterpart of the zero-point
contribution removed by normal ordering in the quantum theory. For the quadratic model this
floor is comparable to the produced entropy and must be subtracted, which is why we report the
produced entropy $\Delta S=S(t)-S(0)$; for the quartic model it is utterly negligible,
$S(0)/\Delta S\sim3\times10^{-4}$; and for the tachyonic model the self-calibration on the
initial vacuum (equivalent to the regularization of Ref.~\cite{GarciaBellido:2001cb}) removes
it exactly, so that $n_k(0)=0$ by construction. In all cases the dynamically produced entropy
is dominated by the highly occupied infrared modes and is insensitive to the ultraviolet.

A second caveat concerns the classical approximation itself. The large occupation numbers that
develop place the system deep in the classical regime $n_k\gg1$ where the classical-statistical
description is justified; the flip side is that classical evolution drives the modes toward a
Rayleigh--Jeans distribution rather than the Bose--Einstein form, so the deep-infrared
occupation numbers, and hence the absolute normalization of $S$, retain a residual dependence
on the lattice cutoff and on the classical approximation. Finally, Eq.~\eqref{eq:smode} is a
single-mode, Gaussian measure that neglects the mode--mode correlations built up in the
nonlinear stage; it therefore provides a lower bound on the true entanglement entropy and is
best interpreted as a measure of the occupation-number irreversibility of preheating.

Finally, it is worth placing these numbers in their cosmological context. The phase-space
entropy computed here is a diagnostic of the nonlinear, far-from-equilibrium preheating stage
within a small comoving patch (the simulation box) and the values $S\sim10^{4}-10^{6}$
count the occupied field modes of that patch, not the eventual entropy of the universe.
Complete reheating requires the inflaton to decay entirely, predominantly through
\emph{perturbative} channels into fermions~\cite{Greene:1998nh} (for which Pauli blocking precludes the parametric
or tachyonic amplification studied here) whose decay products then rescatter and thermalize,
driving the universe into local thermal equilibrium and the radiation era. The associated
entropy is essentially the number of quanta produced, of order $10^{88}$ in a Hubble patch like
ours, vastly larger than anything a lattice box can hold and generated by physics (fermionic
decay and thermalization) beyond the classical, bosonic description used here. Preheating is
thus only the opening, and most violently out-of-equilibrium, act of reheating: it is
precisely this stage, with its large non-adiabatic heat, that can source an appreciable GREA
entropic force, whereas the overwhelming bulk of the final cosmological entropy is laid down
afterwards, as the fluid relaxes into thermal equilibrium. This
raises a natural question for the GREA mechanism itself: if perturbative decay produces the
overwhelming majority of the cosmic entropy, does it not drive a correspondingly large entropic
acceleration? The answer is no. The
entropic force responds not to the total entropy but to the \emph{rate} at which it is produced
relative to the expansion, encoded in $\mathcal{R}=(T\dot S/a^3)/[H(\rho+3p)]$. During
perturbative reheating, the inflaton, oscillating as pressureless matter
($\rho+3p\simeq\rho_\phi$), converts its energy into thermal radiation at the decay rate
$\Gamma$, so the heat density rate is $T\dot S/a^3\simeq\Gamma\rho_\phi$ and thus
\begin{equation}
\mathcal{R}\simeq\frac{\Gamma}{H}\lesssim1 ,
\end{equation}
since reheating completes only when $H(t)$ falls below $\Gamma$. With $\mathcal{R}\le1$ the entropic
term never overcomes the gravitational deceleration,
$q_{\rm GREA}=q_{\rm std}(1-\mathcal{R})\ge0$: the universe decelerates throughout reheating, and the
perturbative decay produces \emph{no} genuine acceleration. Integrated over the roughly one
$e$-fold of the matter-to-radiation transition, the entropic force yields at most a fractional
extra expansion, $\Delta N\lesssim\Gamma/H\sim\mathcal{O}(0.1)$. This is the mirror image of the
explosive preheating burst: perturbative reheating generates an enormous entropy
($\sim10^{88}$) but slowly and quasi-adiabatically ($\Gamma\ll H$ for most of the history,
$\mathcal{R}\ll1$), whereas tachyonic or turbulent preheating dumps a far smaller entropy in
much less than a Hubble time ($\mathcal{R}\gg1$). It is the latter, not the former, that
can source an appreciable ``second burst'' of entropic acceleration. In
any case, once local thermal equilibrium is established, the comoving entropy is conserved,
$\dot S=0$, and this bulk entropic force switches off entirely: the radiation era then expands
adiabatically and decelerates in the standard way. A persistent \emph{late-time} acceleration
must therefore be sourced not by the entropy of matter (frozen after thermalization) but by
the continued growth of the causal-horizon (the boundary) entropy, the term that drives the
present-day acceleration in the GREA framework~\cite{GarciaBellido:2021jdr}.

In the cosmological context of this scenario, it is worth relating the comoving simulation box to the causal
scales of the radiation era. Its size is chosen of order the comoving Hubble radius $(aH)^{-1}$
at reheating, roughly one causal patch, the particle horizon of the incipient hot universe.
That patch is minuscule on cosmological scales: for a reheating scale
$T_{\rm rh}\sim10^{15}\,$GeV, corresponding to
$H_{\rm rh}\simeq1.66\sqrt{g_*}\,T_{\rm rh}^2/M_{\rm Pl}\sim10^{12}\,$GeV, the comoving Hubble
radius has since grown by $a_{\rm rh}H_{\rm rh}/a_0H_0\sim8\times10^{25}$ (using entropy
conservation since then, $a_{\rm rh}/a_0\sim7.8\times10^{-29}$), so the present Hubble volume contains
\begin{equation}
N_{\rm patch}\simeq\left(\frac{a_{\rm rh}H_{\rm rh}}{a_0H_0}\right)^{3}\sim10^{78}
\end{equation}
such boxes. The phase-space entropy $S\sim10^{4}$--$10^{6}$ obtained here is thus that of a
\emph{single} reheating patch; multiplied by the $\sim10^{78}$ patches in our observable
universe, it amounts to $\sim10^{82}$--$10^{84}$, still far below the eventual thermal entropy of
the visible universe, $S_0\sim10^{88}$, the remainder being laid down as the
(predominantly perturbative) decay products rescatter and thermalize into the radiation era.

\section{Conclusions}
\label{sec:conclusions}

We have computed the coarse-grained phase-space entropy produced during preheating from
first-principles lattice simulations with \textsc{CosmoLattice}, and used it to compare the
irreversibility of three canonical reheating scenarios. For the quadratic $m^2\phi^2$ model,
broad parametric resonance drives the daughter occupation numbers to $n_k\simeq7.5\times10^{2}$;
the produced entropy grows by a factor $\simeq2.5$, resides almost entirely in the daughter
sector, and saturates as backreaction ends the resonance. For the quartic $\lambda\phi^4$
model, the occupation numbers cascade to the infrared and exceed $n_k\sim10^{11}$, the produced
entropy grows by more than three orders of magnitude to $\Delta S\simeq4\times10^6$ per
comoving box, and rescattering shares it almost equally between the fields while the system
evolves through a turbulent stage.

For the symmetry-breaking model of Ref.~\cite{GarciaBellido:2001cb} we have shown that the
entropy production admits a closed-form description. The tachyonic growth of the
long-wavelength modes, $n_k+\tfrac12=\tfrac12 e^{2t\sqrt{m^2-k^2}}$, makes the Higgs entropy
grow linearly in time, $S_\phi(t)\simeq(m^4/16\pi)Vt$, until it saturates at symmetry breaking
at $S_\phi^{\rm max}\simeq(m^3V/32\pi)\ln(32\pi^2/\lambda)$. \textsc{CosmoLattice} simulations
of the model reproduce the tanh background of Ref.~\cite{GarciaBellido:2001cb} with
$mt_*\simeq7.9$, the tachyonic dispersion relation $\gamma_k=\sqrt{m^2-k^2}$ across the
unstable band, and the predicted linear entropy law, validating the analytical result. 

These results establish the phase-space entropy as a useful, physically transparent probe of
preheating that complements the usual energetic and spectral analyses and distinguishes its
resonance-dominated, turbulence-dominated, and spinodal regimes. Natural extensions include
mapping the entropy as a function of the resonance parameter and of the coupling $\alpha$,
computing the constant daughter entropy $S_\chi$ from the Bogoliubov spectrum
$n_k^B(\alpha)$ in closed form, following the slow approach to thermal equilibrium at late
times, and extending the analysis to gauge preheating and to multifield sectors.

Finally, we have used this entropy production as the source of the general relativistic
entropic acceleration (GREA) of Refs.~\cite{EspinosaPortales:2021cac,GarciaBellido:2021jdr},
modifying the self-consistent scale-factor evolution of \textsc{CosmoLattice} to include the
entropic force $\tfrac{4\pi G}{3}T\dot S/(a^3H)$, with $T\dot S=\sum_k\omega_k\dot n_k$
reconstructed from the lattice. Applying it to all three models shows that the acceleration is
governed by the \emph{heat} ({\it i.e.} the energy actually transferred to particles) and not by the
phase-space entropy: tachyonic symmetry breaking, which dumps the false-vacuum energy wholesale
into particles in less than a Hubble time, drives a pronounced burst ($q_{\rm GREA}<0$), and the
fragmenting quartic model likewise gives a substantial effect, whereas resonant $m^2\phi^2$
preheating, which produces copious entropy but transfers negligible energy, yields no
appreciable force. The subsequent \emph{perturbative} decay generates the overwhelming majority
of the cosmic entropy ($\sim10^{88}$ quanta in our Hubble patch) but does so slowly and
quasi-adiabatically ($\mathcal{R}\simeq\Gamma/H\lesssim1$), so it drives at most a fractional
extra expansion and switches off once thermal equilibrium freezes the entropy, leaving any
late-time acceleration to the growth of the causal-horizon entropy. A fully self-consistent,
per-step coupling of the entropic force to the field dynamics is a natural next step. All the modifications to \textsc{CosmoLattice} needed to reproduce these results (the model files, the occupation-number output, and the GREA scale-factor kernel) are collected in Appendix~\ref{app:code}.

\

\begin{acknowledgments}
The simulations were performed with the publicly available \textsc{CosmoLattice}
code~\cite{Figueroa:2020rrl,Figueroa:2021yhd}. The author thanks Dani Figueroa for comments on the manuscript and acknowledges support from the Agencia Estatal de Investigaci\'on, under Research Project PID2024-159420NB-C43 [MICINN-FEDER] and the Centro de Excelencia Severo Ochoa Program CEX2020-001007-S at IFT.
\end{acknowledgments}

\appendix

\section{Implementation in CosmoLattice}
\label{app:code}

Here we document the modifications to \textsc{CosmoLattice}~\cite{Figueroa:2020rrl,Figueroa:2021yhd}
required to (i)~simulate the three models, (ii)~output the mode occupation numbers used to build
the phase-space entropy, and (iii)~evolve the scale factor with the GREA entropic force. The code
is organized around user-supplied \emph{model} header files plus a short, model-independent patch
to the scale-factor kernel; the entropy and the heat $T\dot S$ are then assembled from the
standard spectra and volume-averaged energies.

\subsection{Model files and rescalings}
\label{app:models}

The quartic $\lambda\phi^4$ case is the built-in \texttt{lphi4} model. The quadratic and
symmetry-breaking cases are custom models with $N_{\rm s}=2$ scalars (inflaton $\phi$, daughter
$\chi$) and two potential terms. In program variables $\tilde\phi=\phi/f_\star$ the rescalings
$(\alpha,f_\star,\omega_\star)$ are set in the constructor and the potential, its first and second
field derivatives are member functions. For $m^2\phi^2$ (with $q\equiv g^2\phi_0^2/m^2=4q_{\rm res}$):

\begin{widetext}
\begin{verbatim}
    // ---- mphi2.h :  V = 1/2 phi~^2  +  1/2 q phi~^2 chi~^2 ----
    m = parser.get<double>("mass");
    q = parser.get<double>("q");                  // q = g^2 phi0^2 / m^2
    fldS0 = parser.get<double,2>("initial_amplitudes");
    piS0  = parser.get<double,2>("initial_momenta", {0,0});
    alpha = 0;   fStar = fldS0[0];   omegaStar = m;
    setInitialPotentialAndMassesFromPotential();

    auto potentialTerms(Tag<0>) { return 0.5   * pow<2>(fldS(0_c)); }
    auto potentialTerms(Tag<1>) { return 0.5*q * pow<2>(fldS(0_c)*fldS(1_c)); }
    auto potDeriv (Tag<0>) { return fldS(0_c) + q*fldS(0_c)*pow<2>(fldS(1_c)); }
    auto potDeriv (Tag<1>) { return             q*fldS(1_c)*pow<2>(fldS(0_c)); }
    auto potDeriv2(Tag<0>) { return 1.0       + q*pow<2>(fldS(1_c)); }
    auto potDeriv2(Tag<1>) { return             q*pow<2>(fldS(0_c)); }
\end{verbatim}
\end{widetext}

The symmetry-breaking model uses the double well
$\tilde V=\tfrac14(\tilde\phi^2-1)^2+\tfrac12 q\,\tilde\phi^2\tilde\chi^2$ ($q=g^2/\lambda$),
$f_\star=v$, $\omega_\star=m=\sqrt{\lambda}\,v$, $\alpha=0$, and the expansion switched off ($a=1$)
for the fast quench:

\begin{widetext}
\begin{verbatim}
    // ---- ssb.h :  V = 1/4 (phi~^2 - 1)^2  +  1/2 q phi~^2 chi~^2 ----
    lambda = parser.get<double>("lambda");
    v      = parser.get<double>("v");
    q      = parser.get<double>("q");             // q = g^2 / lambda
    fldS0  = {0.0, 0.0};                          // phi starts at the top
    alpha  = 0;   fStar = v;   omegaStar = sqrt(lambda)*v;
    setInitialPotentialAndMassesFromPotential();

    auto potentialTerms(Tag<0>) { return 0.25*pow<2>(pow<2>(fldS(0_c)) - 1.0); }
    auto potentialTerms(Tag<1>) { return 0.5*q*pow<2>(fldS(0_c)*fldS(1_c)); }
    auto potDeriv (Tag<0>) { return pow<3>(fldS(0_c)) - fldS(0_c)
                                  + q*fldS(0_c)*pow<2>(fldS(1_c)); }
    auto potDeriv (Tag<1>) { return   q*fldS(1_c)*pow<2>(fldS(0_c)); }
    auto potDeriv2(Tag<0>) { return 3.0*pow<2>(fldS(0_c)) - 1.0
                                  + q*pow<2>(fldS(1_c)); }
    auto potDeriv2(Tag<1>) { return   q*pow<2>(fldS(0_c)); }
\end{verbatim}
\end{widetext}

Because $\phi$ starts at the maximum, $V''(0)=-1<0$ in program units. \textsc{CosmoLattice} detects
the negative initial mass squared and seeds that field with the massless vacuum dispersion
$\omega_k=|k|$, i.e.\ $\phi_k(0)=1/\sqrt{2k}$, exactly the initial condition of
Ref.~\cite{GarciaBellido:2001cb}; no further intervention is needed to trigger the spinodal growth.

\subsection{Occupation numbers and the phase-space entropy}
\label{app:entropy}

Setting \verb+OccNumb = true+ in the input file turns on the per-field occupation-number output:
for each scalar the code writes, at every spectral snapshot, a radially binned spectrum with columns
$\{k,\;P_\phi=\langle|\tilde\phi_k|^2\rangle,\;P_\pi=\langle|\tilde\pi_k|^2\rangle,\;n_k,\;
\mathcal{N}_{\rm bin}\}$, where $\mathcal{N}_{\rm bin}$ counts the lattice modes in the shell. The
phase-space entropy [Eq.~\eqref{eq:Stot}] is assembled in post-processing as
$S=\sum_{\rm bins}\mathcal{N}_{\rm bin}\,s(n_{\rm bin})$, $s(n)=(1+n)\ln(1+n)-n\ln n$.

For the parametric models we use the built-in estimator $n_k$ directly. In the tachyonic model it
is unreliable, since it references $\omega_k=\sqrt{k^2+a^2m_{\rm eff}^2}$, imaginary while
$m_{\rm eff}^2<0$; we instead reconstruct the frequency-independent occupation number of
Ref.~\cite{GarciaBellido:2001cb}, $n_k+\tfrac12=|\tilde\phi_k\tilde\pi_k|=\sqrt{P_\phi P_\pi}$,
self-calibrated on the initial vacuum so that $n_k(0)=0$,
\begin{equation}
n_k(t)=\tfrac12\!\left[\frac{\sqrt{P_\phi(k,t)\,P_\pi(k,t)}}{\sqrt{P_\phi(k,0)\,P_\pi(k,0)}}-1\right],
\label{eq:nk_reconstruct}
\end{equation}
equivalent to the regularized dispersion of Ref.~\cite{GarciaBellido:2001cb} and removing the
ultraviolet vacuum floor by construction.

\subsection{The heat source $T\dot S$}
\label{app:heat}

The GREA force is sourced by the heat $T\dot S$. Assigning each mode the bosonic temperature
$T_k=\omega_k/\ln(1+1/n_k)$ and using $ds_k/dn_k=\ln(1+1/n_k)$, the logarithms cancel and
$T\dot S=\sum_k T_k\dot s_k=\sum_k\omega_k\dot n_k$. On the lattice we evaluate this as the
non-adiabatic energy transfer into the incoherent (kinetic$+$gradient) sector,
\begin{equation}
\frac{T\dot S}{a^3}=\dot\rho_{\rm part}+3H(\rho_{\rm part}+p_{\rm part}),
\label{eq:heat_bookkeeping}
\end{equation}
splitting off the homogeneous condensate,
$\rho_{\rm cond}=\tfrac12 a^{-2\alpha}\langle\tilde\phi'\rangle^2+V(\langle\tilde\phi\rangle)$
(with the $a^{-2\alpha}$ factor arising from the momentum rescaling $\pi=a^{3-\alpha}\phi'$), from
the fluctuations, $\rho_{\rm part}=\rho-\rho_{\rm cond}$. Both the volume-averaged energies and the
homogeneous means $\langle\tilde\phi\rangle,\langle\tilde\phi'\rangle$ are standard outputs.

\subsection{GREA scale-factor evolution}
\label{app:kernel}

The self-consistent expansion integrates the program-variable acceleration equation through a
scale-factor kernel. Two public members are added to the base model class
(\texttt{abstractmodel.h}),
\begin{widetext}
\begin{verbatim}
    bool greaOn  = false;   // enable the GREA entropic force
    T    TdSrate = 0;       // heat rate T dS/dt (program units), set by the measurer
\end{verbatim}
\end{widetext}
and the kernel (\texttt{scalefactorkernels.h}) is augmented with the entropic
term $\tfrac{4\pi G}{3}\,T\dot S/(a^3H)$ of Eq.~\eqref{eq:grea3}:
\begin{widetext}
\begin{verbatim}
    auto standard =
        pow(model.aI, 2*model.alpha+1)/3.0 * pow<2>(model.fStar/Model::MPl)
      * ( (model.alpha-2)*(Eks + Ekcs + EkSU2Dbl)
        + model.alpha    *(Egs + Egcs + EgSU2Dbl)
        + (model.alpha-1)*(EelU1 + EmagU1 + EelSU2 + EmagSU2)
        + (model.alpha+1)* model.potAvI );

    if (model.greaOn) {                 // GREA entropic force  (alpha = 0)
        auto H = model.aDotI / model.aI;
        standard += pow(model.aI, 2*model.alpha+1)/3.0
                  * pow<2>(model.fStar/Model::MPl)
                  * 0.5 * model.TdSrate / (pow<3>(model.aI) * H);
    }
    return standard;
\end{verbatim}
\end{widetext}
The flag \verb+greaOn+ is read from the input, and \verb+TdSrate+$=T\dot S$ is updated by the
measurer at the measurement cadence (an operator split). With the entropic term active the
Hamiltonian constraint~\eqref{eq:grea2} is monitored but no longer exactly enforced, its violation
quantifying the entropic energy exchange. For the results of Sec.~\ref{sec:grea} the heat is
evaluated as in Eq.~\eqref{eq:heat_bookkeeping} and the coupled Friedmann system integrated with
it; a fully in-loop, per-step evaluation of \verb+TdSrate+ from the instantaneous spectra is the
natural extension.

\subsection{Run parameters}
\label{app:params}

A representative input file (symmetry-breaking model) reads
\begin{widetext}
\begin{verbatim}
    expansion = true        evolver = VV2
    N = 128    dt = 0.02    kIR = 0.02        # 128^3, box L = 100 pi / m
    OccNumb = true          tOutputInfreq = 0.5
    kCutOff = 2.5
    initial_amplitudes = 0 0      initial_momenta = 0 0
    lambda = 1e-4    v = 1e16    q = 2        # q = g^2 / lambda = 4 q_res
\end{verbatim}
\end{widetext}
Compilation selects the model via \verb+cmake -DMODEL=mphi2 ..+ (or \verb+lphi4+, \verb+ssb+)
followed by \verb+make cosmolattice+, with \textsc{fftw3} as the only external dependency.

\bibliographystyle{apsrev4-2}
\bibliography{refs_preheating}

\end{document}